\documentclass[lettersize,journal]{IEEEtran}
\usepackage{amsmath,amsfonts}
\usepackage{algorithmic}
\usepackage{algorithm}
\usepackage{array}
\usepackage[caption=false,font=normalsize,labelfont=sf,textfont=sf]{subfig}
\usepackage{textcomp}
\usepackage{stfloats}
\usepackage{url}
\usepackage{verbatim}
\usepackage{graphicx}
\usepackage{cite}
\usepackage[normalem]{ulem}
\usepackage{xcolor}

\newcommand{\shaolun}[1]{\textcolor{black}{#1}}

\graphicspath{{figs/}{figures/}{pictures/}{images/}{./}{photos/}}
\newcommand{\toolName}[1]{\textit{StoryExplorer}}
\begin{document}

% --------------------------------------------------------
\title{\toolName{}: A Visualization Framework for Storyline Generation of Textual Narratives}

\author{
Li Ye, Lei Wang, Shaolun Ruan, Yuwei Meng*, Yigang Wang, Wei Chen and Zhiguang Zhou*
        % <-this % stops a space
\thanks{Li Ye and Lei Wang contribute equally to this work.}
\thanks{Li Ye, Lei Wang, Yigang Wang and Zhiguang Zhou are with Hangzhou Dianzi University. E-mail: \{liye, leiwang, yigang.wang, zhgzhou\}@hdu.edu.cn}
\thanks{Shaolun Ruan is with Singapore Management University. E-mail: slruan.2021@phdcs.smu.edu.sg}
\thanks{Yuwei Meng is with Zhejiang Provincial Energy Group Company Ltd. E-mail: mengyuwei@zhenergy.com.cn}
% \thanks{Wei Chen is with Singapore Management University. E-mail: slruan.2021@phdcs.smu.edu.sg}
% \thanks{Yongheng Wang is with Zhejiang Lab. E-mail: wangyh@zhejianglab.com}
\thanks{Wei Chen is with Zhejiang University. E-mail: chenvis@zju.edu.cn}
% \thanks{Yong Wang is with Singapore Management University. E-mail: yongwang@smu.edu.sg}
\thanks{(*Corresponding author: Yuwei Meng and Zhiguang Zhou)}
}

\markboth{Journal of \LaTeX\ Class Files,~Vol.~14, No.~8, August~2021}%
{Shell \MakeLowercase{\textit{et al.}}: A Sample Article Using IEEEtran.cls for IEEE Journals}


\maketitle
%-------------------------------------------------------------------------
\begin{abstract}
\shaolun{In the context of the exponentially increasing volume of narrative texts such as novels and news, readers struggle to extract and consistently remember storyline from these intricate texts due to the constraints of human working memory and attention span.}
To tackle this issue,
\shaolun{we propose a visualization approach \toolName{}, which facilitates the process of knowledge externalization of narrative texts and further makes the form of mental models more coherent.}
% a novel storyline generation workflow to , 
\shaolun{Through the formative study and close collaboration with 2 domain experts}, 
we identified key challenges for the extraction of the storyline. 
\shaolun{Guided by the distilled requirements, we then propose a set of workflow (i.e., insight finding-scripting-storytelling) to enable users to interactively generate fragments of narrative structures.}
We then propose a visualization system \toolName{} which combines stroke annotation and GPT-based visual hints to quickly extract story fragments and interactively construct storyline.
% \shaolun{xxx. \shaoluncomment{@Yeli, please use one sentence to highlight the issue our vis tool gonna solve and fill the xxx here.}}
\shaolun{To evaluate the effectiveness and usefulness of \toolName{},
we conducted 2 case studies and in-depth user interviews with \textcolor{black}{16} target users. 
The result shows that users can better extract the storyline by using \toolName{} along with the proposed workflow.
}


\end{abstract}  
%-------------------------------------------------------------------------
\vspace{-5mm}
\section{Introduction}
\IEEEPARstart{N}{arrative} structure is a commonly used abstraction in understanding and interpreting various entities, particularly in the realm of textual narratives such as novels and news.
This type of structure, essentially a framework for storytelling, plays a crucial role in how information is organized and conveyed,
\shaolun{which can significantly impact the reader's comprehension and retention~\cite{barthes1975introduction}.}
\shaolun{Notably, the importance of narrative structure becomes even more pronounced in the digital age,}
where massive narrative text volumes create information overload and challenge cognitive processing~\cite{richgels1987awareness}.
% Extracting and understanding narrative structures from complex texts is crucial, not only for literary analysis but also for broader applications in education, journalism, and data visualization.

However, \shaolun{with the volume of textual contents significantly increasing (e.g., textual data, the number of entities, and the complexity of events), it becomes increasingly difficult}
for readers to extract narrative structures from textual data and acquire persistent memory~\cite{min2019modeling}.
\shaolun{Specifically}, according to the process of working with the domain expert and literature review~\cite{Kintsch_van_Dijk_2006,Chi_2009}, 
the limited working memory, attention, and other cognitive resources~\cite{richgels1987awareness,cote1998students} 
% gradually become depleted during the reading process,
significantly increasing the readers' cognitive load,
making them struggle to form a mental structure of the narrative texts.
%  leading to the problem of cognitive overload. 
\shaolun{Also, important entities or events may be missed or forgotten}~\cite{kormos2011task}, which may cause comprehension difficulties and reduce reading efficiency.

To enhance the process of the perception of textual narrative structure, 
a straightforward solution is to externalize important information (e.g., people and events) as well as other features of the narrative structure implicit in the text,
% which can conserve mental resources and achieving the objectives of cognitive improvement and effective text comprehension. 
\shaolun{To achieve this, one common approach is to employ NLP models, such as BERT~\cite{devlin2018bert} and BART~\cite{lewis2019bart}}.
\shaolun{However, they have limitations in extracting coherent story plots from lengthy texts and thus often produce outputs that lack intuitive understanding and personalization.}
This gap underscores the need for narrative visualization as an effective strategy for knowledge externalization, addressing the shortcomings of current text-mining techniques in capturing the essence of complex narrative structures.
Visualization has proven to be a good way to support the process of knowledge externalization~\cite{burkhard2005knowledge}.

To this end, we present \toolName{}, 
\shaolun{a visualization tool to generate storyline from narrative texts based on knowledge externalization.}
First, 
\shaolun{we propose a workflow to represent the process of narrative text knowledge externalization}
to better support user iteratively generate storyline from raw texts.
Building upon the workflow, we develop \toolName{},
\shaolun{which assists users in locating important elements in the narratives.}
\toolName{} consists of three main views: Text View, Fragment View, and Storyline View, which correspond to the three main stages of the proposed workflow ``insight finding-scripting-storytelling''. Specifically, Text View extracts basic narrative elements through stroke annotations and GPT-based visual hints. Fragment View organizes these elements to help users quickly edit basic narrative units. Storyline View serves as the final stage of knowledge externalization, supporting rapid editing, browsing and sharing of the final storyline to facilitate rapid iteration.
\shaolun{We demonstrate our approach by case studies and in-depth user interviews, and the results show that} 
\toolName{} can better externalize knowledge from narrative texts while facilitating the generation of mental models.
% Figure~\ref{Fig1} illustrates a simple storyline generation process.
Our contributions are summarized as follows:
% \vspace{-1mm}
\begin{itemize}
    \setlength{\itemsep}{0pt}
	\setlength{\parsep}{0pt}
	\setlength{\parskip}{0pt}
    \item
    % Representing the process of narrative text knowledge externalization by extending the selection-organization-integration process of active reading to insight finding-scripting-storytelling.
    We generalize the workflow of narrative text knowledge externalization to insight finding-scripting-storytelling, which can enable users to quickly generate storyline from narrative texts.
    % Users begin by identifying various narrative elements within the text, which can be further organized into expressive fragments and ultimately integrated into a coherent storyline.
    % \item \textbf{Human-Computer integration approach for story fragment generation.} A WYSIWYG story fragment generation method that decomposes the user's narrative process into basic entities and their relationships, abstracts the story fragment into a quadruplet of persons, time, place and events, and uses natural language processing techniques to recommend the narrative process.
    \item
    We propose \toolName{}, an interactive visualization system that supports our workflow to help users gain a better understanding of the narrative texts while facilitating the generation of mental models. 
    % The system integrates stroke annotation and GPT-based visual hints, enhancing the iterative development of the fragment and providing interactive assistance for exploring and modifying the storyline.

    \item \shaolun{We conducted careful-designed case studies and in-depth user interviews to evaluate the effectiveness and usefulness of our proposed visualization tool \toolName{}.}
\end{itemize}

\begin{figure*}[tb]
 \centering
 \includegraphics[width=0.85\linewidth]{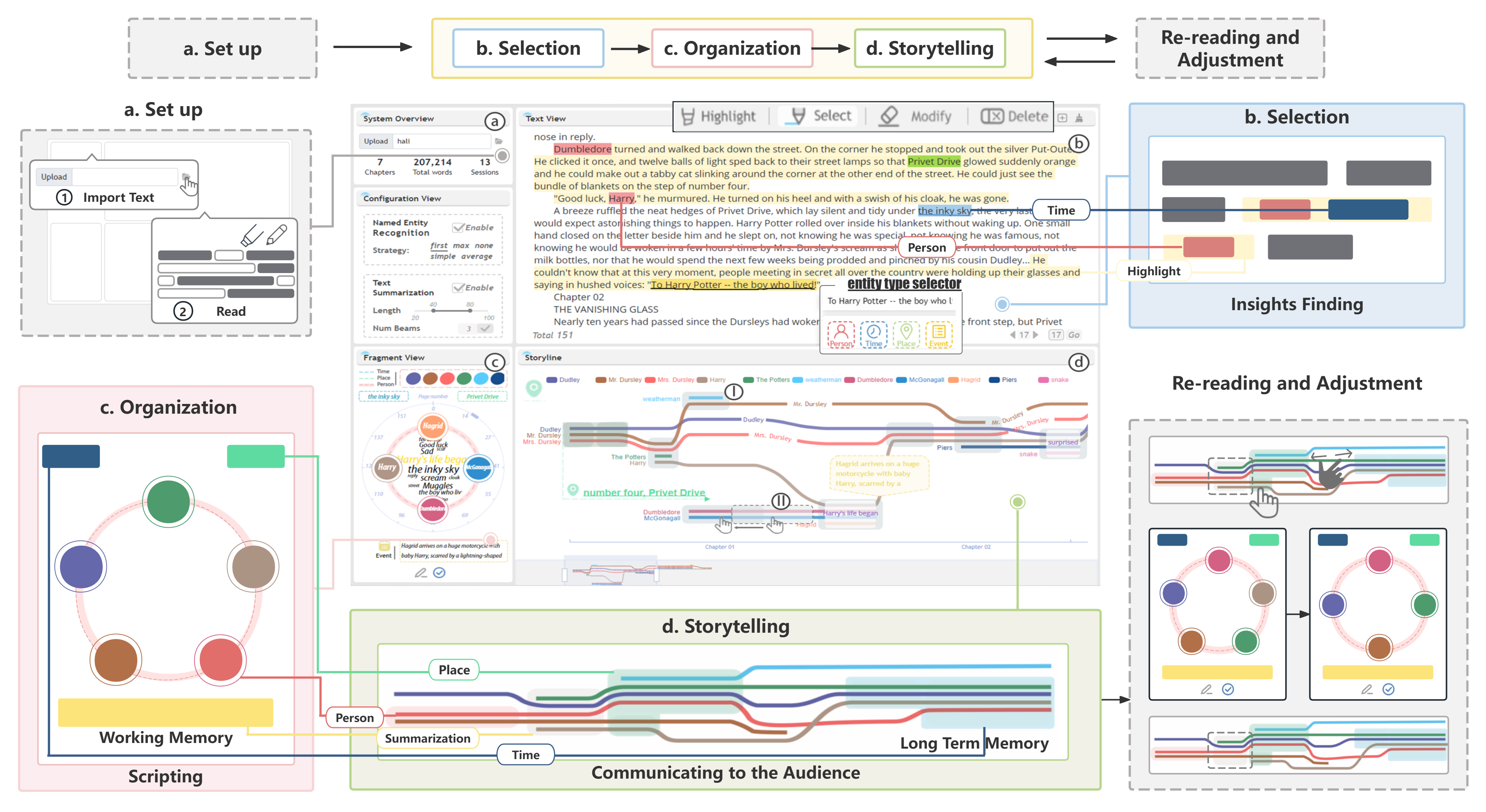}
  \caption{\textcolor{black}{The workflow of narrative text knowledge externalization using \toolName{}. 
  % \shaoluncomment{@Yeli, please modify the teaser according to our discussion on WeChat.} 
  (b)\toolName{} allows readers to highlight key sentences in the text, and then \toolName{} recognizes entities using GPT-based visual hints, followed by the selection of entities by stroke annotation.
  (c)Readers can organize these entities, and then add additional information, i.e., keywords and event summarization, to form fragments.
  (d) 
  % Finally, multiple fragments are integrated into a storyline. 
  Readers can further explore and modify the storyline to create a unified storytelling representation.
  }}
  \vspace{-3mm}
\label{Fig1}
\end{figure*}

\section{RELATED WORK}
We categorized the prior work related to storyline generation of textual narratives into two groups, i.e., storyline visualization and knowledge externalization strategies.
% In this section, we review the recent studies that are most relevant to our work, including storyline visualization, (narrative texts) knowledge externalization strategies, and natural language processing.

\vspace{-2mm}
\subsection{Storyline Visualization}
\shaolun{Prior studies have explored the design space of storyline layout algorithms for producing artistically-quality storyline visualizations.}
% in a reasonable amount of time. 
\shaolun{This research direction was initially inspired by Munroe's movie narrative charts \cite{movienarrativecharts} titled ``Movie Narrative Charts''.}
% , which depicts the plots of several famous films in terms of character interaction through the intersection and overlap between the line segments.
Kim et al. \cite{kim2010tracing} extended Munroe's movie narrative charts, but they advanced the technique by restricting its applicability to specific data structure.
% \shaoluncomment{which data type? make it specific.}
% , resulting in a more efficient and rapid layout algorithm. 
Ogawa and Ma \cite{ogawa2010software} presented a technique for visualizing the interactions between developers in software project evolution. 
% However, the fact that the heuristic rules are relatively simple prevents the algorithm from producing results as good as those created by professional artists.
Tanahashi and Ma \cite{tanahashi2012design} proposed a set of design considerations and formulated the storyline layout problem as a genetic algorithm-based optimization method. The method can successfully produce an aesthetically pleasing and legible storyline layout that is comparable to those produced by professional artists. 
% However, when the number of characters and time frames reaches hundreds, creating a storyline layout takes considerable time. 
Arendt and Pirrung~\cite{arendt2017matters} proposed a y-axis coordinate environment consistency method, by encoding interaction context to have a consistent interpretation over time.
More recent research has focused on the fast and automatic construction of storyline layouts \cite{tang2018istoryline,tang2020plotthread,hulstein2022geo,deng2023visualizing}. These approaches use more effective deep learning-based natural language processing algorithms and human-computer interaction processes to facilitate storyline layout adjustment and optimization.
% but this is not suitable for web-based visualization. In practice, this approach is resource-intensive and less efficient.

However, existing work on the storyline focused on storyline layout adjustment and style optimization but did not focus on abstracting the process of organizing raw text into \textcolor{black}{the} storyline. 
\textcolor{black}{While Text2Storyline\cite{gonccalves2023text2storyline} offers automatic storyline generation through explicit temporal information extraction, \toolName{} provides a more flexible human-in-the-loop approach that combines stroke annotation and GPT-based visual hints to capture both explicit and implicit temporal relationships in narrative texts.}

\vspace{-2mm}
\subsection{Knowledge Externalization Strategies}
\shaolun{Prior study has pointed out that rising text volumes and increasing complexity can challenge readers' ability to convert working memory into long-term retention during reading~\cite{richgels1987awareness}.}
% , as identified by Richgels . 
This issue may lead to inefficient reading practices, hindering readers from grasping the text's context and narrative. 
However, \shaolun{methods like active reading, note-taking, and graphical modeling \cite{cox1999representation, chi2009active} have been proposed to facilitate the creation of coherent mental models~\cite{kintsch1988role}. For example, }
% , offering lasting and editable comprehension aids.
% \shaoluncomment{@Yeli, please rewrite this paragraph to make it more clear: 1) give more examples; 2) illustrate each paper in short.}
the construction-integration model \cite{kintschcbemafrs} describes reading as a process where readers initially absorb key text concepts into working memory, then organize these into fragmented, sentence-level pieces, and finally integrate them into coherent expressions using specific rules. However, as the working memory load increases, comprehension and attention may decline. To counteract this, research has explored enhancing reading efficiency with external aids \cite{cox1999representation, scaife1996external}.

\shaolun{Our work} draws inspiration from active diagramming tools based on pen-and-ink interactions \cite{subramonyam2020texsketch} to provide solutions for \toolName{}. One approach is to integrate different diagram elements through appropriate interactions by means of pre-defined layout rules \cite{lu2018inkplanner}. The other focuses on the automatic conversion of the extracted paradigms into geometric and graphical representations by logical representation \cite{seo2014diagram}. 
% \shaoluncomment{I'm still not clear about the relationship of the above three papers and the necessity of introducing them here.}
\toolName{} combines both approaches, automatically converting selected text into entity units and creating entity relationships directly through panel interactions. To automatically recognize entities, we also employ a large language model. 

\begin{figure*}[tb]
    \centering
    \includegraphics[width=\linewidth]{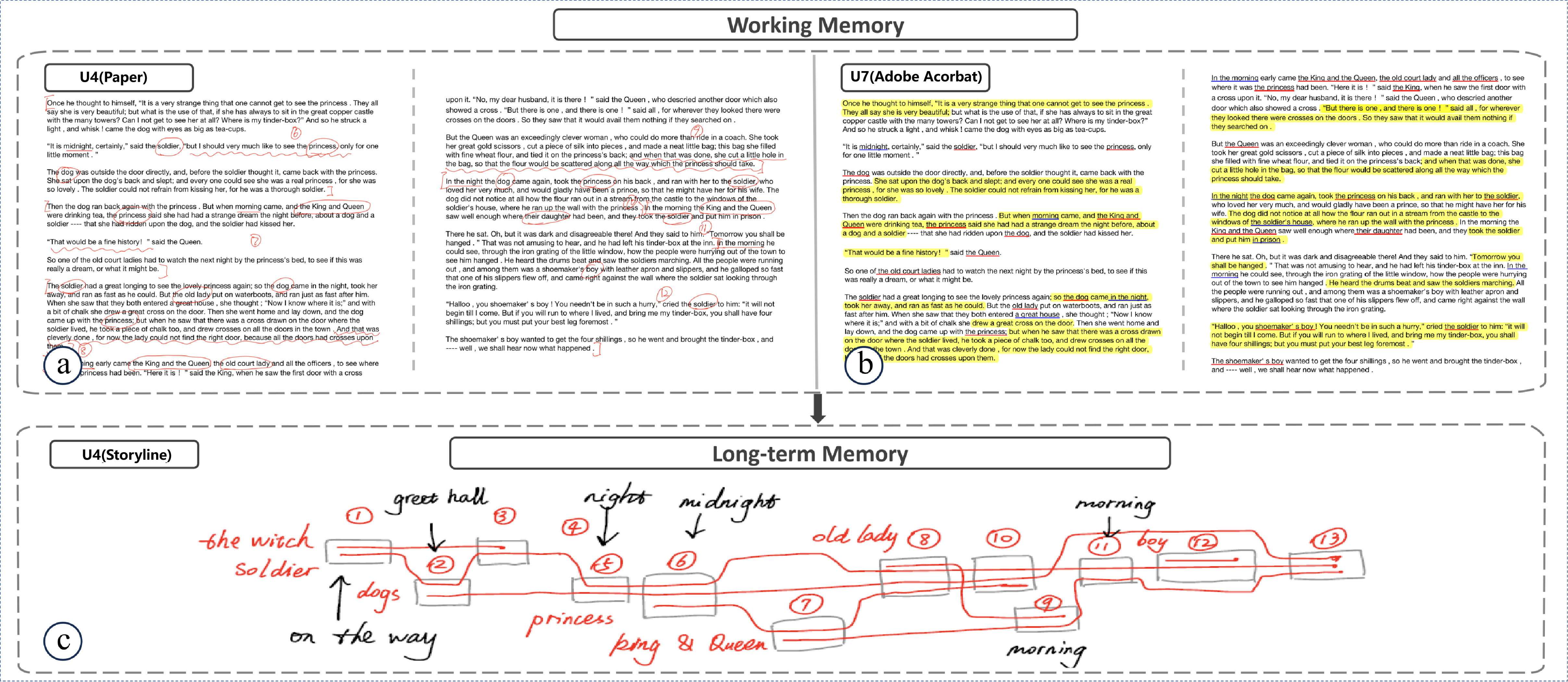}
    \caption{\textcolor{black}{Formative study result: Participants annotated raw text from (a) entity books and (b) Adobe Acrobat to form working memory and ultimately conveyed the working memory using (c) storyline to promote long-term memory.}}
    \label{Fig2}
    \vspace{-6mm}
\end{figure*}

\section{FORMATIVE STUDY}

\shaolun{We conducted a well-designed preliminary study to collect the actual need that could aid in extracting narrative structure and developing a more accurate model of readers' behavior.}
\shaolun{To guarantee our prototype could seamlessly fit into the readers' reading habits, we devised a comparison study for investigating the reading behavior of creating a storyline. }
That is, the reader extracts the narrative structure using an entity book and text-reading tools (such as Adobe Acrobat) and then draws the storyline on paper based on the extraction.

\subsection{Procedure}
\textcolor{black}{We recruited 16 participants (8 undergraduate and 8 graduate students, U1-U16) from a local university, representing disciplines in Computer Science, English, and Art Design. All participants demonstrated proficiency in English reading comprehension and visual note-taking, though none had prior storyline creation experience.
The experiment centered on extracting fragments from the narrative text "The Tinder-Box" using conventional tools (entity books and Adobe Acrobat). While no strict time constraints were imposed, task completion duration was recorded as a performance metric. Participants received compensation of \$10 upon study completion.}

First, we described the elements required to generate a storyline and how the storyline was generated. The reading of the text commenced once the participants indicated comprehension. Participants were provided with A4-sized paper reading materials and several blank sheets of paper to draw the storyline. The participants could also use different colored pens to highlight distinct elements. The participants could annotate the text on a PC using Adobe Acrobat(Fig.~\ref{Fig2}). All participants participated in a semi-structured interview at the end of the study to provide feedback, which was recorded on video for later analysis.

\subsection{Design Considerations.}
We worked closely with two experts (E1 \& E2) to extract the detailed requirements and collect their feedback.
E1 is an expert in human-computer interaction with 5 years of experience in authoring tools.
E2 is a cognitive psychologist with a decade of expertise in researching how humans process and comprehend language.
\textcolor{black}{We analyzed 32 storylines (16 participants x 2 storylines)}, interview notes, and video recordings, and through repeated observations of the video data, the following three design considerations were derived from all aspects of narrative structure extraction to storyline generation.

\noindent
\textbf{DC1. Provide explicit workflow for interactive methods.}
Based on these observations, it is essential to provide a clear workflow for our interactive approach in order to reduce the time wasted in switching workflows.
This system generalizes the select-organize-integrate \cite{chi2009active,mayer1984aids} approach in active reading into insights finding-scripting-storytelling~\cite{lee2015more} approach to represent the storyline construction process in order to generate a rational narrative structure construction process. \textcolor{black}{This generalization aims to align our system with established cognitive processes while reducing cognitive load and increasing efficiency in storytelling.}

\noindent
\textbf{DC2. Enable users to construct fragments rapidly.}
\textcolor{black}{Different academic backgrounds influenced users' text processing strategies: Education majors emphasized pedagogical structure, while English majors focused on narrative elements. To accommodate these diverse approaches, our system combines NLP-based entity suggestions with flexible annotation tools. This hybrid approach allows users to construct fragments efficiently while maintaining their preferred analytical methods.}

\noindent
\textbf{DC3. Create a storyline that is easy to observe and iterate.}
% \shaoluncomment{I would recommend to remove this DC3, or merge it into other requirements.}
\toolName{} should combine and unify the storyline with the timeline, supporting panning and zooming operations while keeping the storyline view visually uncluttered, thus preventing the storyline from being too long to view or the window from being too small to comprehend the overall story. At the same time, each fragment is displayed and modified in real-time via the fragment view, and the modified results are rapidly iterated using incremental compilation. \textcolor{black}{This approach enhances the user's ability to grasp the overall story while allowing for detailed exploration and modification of specific elements.}

\section{STORYEXPLORER}
In this section, we first present a usage scenario to illustrate the workflow of constructing storyline using \toolName{}. We then describe its user interface and the technologies employed in its implementation.

\subsection{Workflow for Storyline Generation}
To understand how \toolName{} supports storyline generation, we follow Clover. Now Clover is about to read~\textbf{Harry Potter and the Sorcerer's Stone} and begin to understand the story. 

\subsubsection{Set up}
Clover initiates the system and uploads the text via the \textcolor{black}{``Upload''} button. Upon uploading, the text is loaded in the text view (Figure~\ref{Fig1}b), the storyline constructed from existing fragments is loaded in the storyline view (Figure~\ref{Fig1}d), and the overview panel (Figure~\ref{Fig1}a) displays the basic information of the system. Afterward, a new fragment will be added to the fragment view (Figure~\ref{Fig1}c), guiding Clover through the fragment organization.

\subsubsection{StoryLine Generation}
Clover starts reading and mentally organizing the fragments, but the fragments produced by one-pass reading are incomplete and unreliable, so the two-pass reading approach is frequently recommended for narrative texts.
% When Clover reads, he highlights the text of entities that may be present with \textit{Highlight} tools, saves them in working memory, and uses this information to organize the fragments in a second reading.
During his reading, Clover uses the \textit{Highlight} tool to mark potential entities, storing them in his working memory for reorganization in the second pass.
In addition, through named entity recognition, the system highlights entities with different background colors in the highlighted text, and automatically adds the identified entities to the fragment panel.
This process minimizes working memory loss and enhances fragment organization efficiency in the second reading (DC2).
% thereby reducing the loss of working memory during the second reading and increasing the efficiency of fragment organization (D2). 
This use of NLP models stems from our formative study, which identified the time-intensive nature of re-locating key fragment elements in subsequent readings. 
% The idea of utilizing NLP models originated from our formative study, which revealed that readers must read the text repeatedly to re-locate the key elements of a fragment in a second reading, which is typically time-consuming.
% Now, Clover has just finished reading a story about Professors Dumbledore and McGonagall leaving Harry on his Muggle relatives' doorstep, and trying to build the fragment backward. With \toolName{}, Clover works to comprehend the text while quickly organizing the fragments to generate a storyline. His workflow involves selection, organization, and storytelling (D1) (see Figure~\ref{Fig3}).
Clover has just finished a story about Professors Dumbledore and McGonagall leaving Harry with his Muggle relatives and is now attempting to reconstruct the fragment retrospectively. With \toolName{}, he efficiently organizes the fragments while understanding the text, following a workflow of selection, organization, and storytelling (DC1).

\noindent
\textbf{Selection (Insights Finding)}
Clover re-read the text, selecting fragment elements based on his preliminary mental organization.
% He will see "Professor Dumbledore" and "Harry" with a red background color and "Privet Drive" with a green background color in the passage. The different colors correspond to the person's colors of different fragment elements in the system according to the color-gestalt, and Clover is free to disregard these NLP-based cues.
In the passage, he notices \textcolor{black}{``Professor Dumbledore''} and \textcolor{black}{``Harry''} highlighted in red, and \textcolor{black}{``Privet Drive''} in green. These colors align with the system's color gestalt for different fragment elements, representing persons and places.
% At this point, Clover seems to be less satisfied with the NLP-based cues, and he modifies the model recognition result (D2) by changing the recognition strategy option for named entity recognition on the operation panel (which also allows him to choose whether to enable the model or not), and determines whether the re-identified entity (i.e., text with a red background color: Dumbledore and Harry) meets the requirements.
Similarly, while prepared to add time elements based on GPT-based hints, Clover finds the identified locations unsatisfactory.
Therefore, \toolName{} also supports underlining entities and automatically recognizing the underlined content through stroke annotation.
Using the \textit{Select} tool, Clover can underline entities for selection.
% If the phrase spans multiple lines, Clover can underline multiple lines of text and the system will automatically recognize the same entity as being added to the fragment. Finally, NLP-identified or stroke annotated entities are automatically added to the fragment view for further organization stage.
If an entity spans multiple lines, he can underline them all, and the system will recognize and add them as a single entity to the fragment. Ultimately, GPT-identified and stroke-annotated entities are automatically incorporated into the fragment view for the next stage of organization.

\noindent
\textbf{Organization (Scripting)}
% After deciding which words or phrases to underline, Clover can classify them into various types of entities. To do this, after Clover underlines "the inky sky", the entity type selector automatically appears, allowing him to select person, time, place, or event and automatically add them to the corresponding position in the fragment view.
After selecting words or phrases to underline, Clover categorizes them into different entity types. To do this, when he underlines \textcolor{black}{``the inky sky''}, an entity type selector pops up, enabling him to classify it as a person, time, place, or event. This classification directly places the entity in the fragment view.
% Similarly, the entity type selector can be modified for a variety of NLP-identified or manually annotated entity types, and the user can modify the entity's content or type using the \textit{Modify} tool at the top of the text view (D2). Then, Clover chose the time option, and "the inky sky" was placed in the upper-left corner of the fragment view.
Additionally, Clover can adjust the type and content using the \textit{Modify} tool (DC2). Then, Clover chose the time option, and \textcolor{black}{``the inky sky"} was placed in the upper-left corner of the fragment view.

% The fragment view serves to display and edit the elements forming the fragment, and entities annotated by NLP or stroke annotation are arranged there. Unlike previous work, we separate the organization operation from the raw text by making the place where it occurs independent of other stages, thereby better decoupling the features of the narrative structure and presenting it dynamically (D1).
The fragment view serves to display and edit the quadruples forming the fragment. Diverging from prior approaches, we isolate the organization process from the raw text, situating it in a distinct stage. This separation enhances the narrative structure's features and dynamic presentation (DC1).
The top of the view provides a scrollable list of person entity alternatives, the outer arc of the center diagram indicates the page range of the current fragment, and the inner arc is composed of person entities, which indicate their relationships. Inside the diagram is a word cloud generated from the fragment's text, and a text block below displays the fragment's overall content.
% We also note that not all fragments have a complete quadratic structure, and some elements may be hard to find in the text, so none of the three elements are required except for the person entity (D2).
Some elements might be elusive in the text, making only the person entity mandatory among the three elements (DC2).

% Currently, Clover has selected persons and place, and the fragment view is simultaneously organized rationally.
Currently, Clover begins to summarize the fragment's content. At this point, Clover is surprised that the event content box is already filled with text. Text summarization is used to briefly summarize the content of the fragment.
% and when the user does not underline the event content in the text, the text summary extraction model automatically updates the content iteratively according to the user's highlighted areas (D2).
When the user doesn't underline event content, this model iteratively updates the summary based on the user's highlighted areas (DC2).
% Clover can also select a number of words that best express the message of the fragment based on the system-generated word cloud. 
% These words will be displayed directly on the storyline fragment, allowing Clover to quickly recall the fragment's general content.
Additionally, Clover can choose keywords from the word cloud that best convey the fragment's message. These words are then displayed on the storyline fragment, aiding Clover in quickly recalling the fragment's general content.
Once all elements are organized, the fragment is finalized and progresses to the storytelling phase.

\noindent
\textbf{Integration (Storytelling)}
% At this stage, Clover integrates several previously organized fragments into a storyline. That is, he integrated the propositional representations into a mental model of the text and constructed a unified understanding.
At this stage, Clover integrates several pre-organized fragments into a coherent storyline.
% In the storyline view, he finds it very convenient that the storyline is updated iteratively as the fragments are organized.
% In other words, we use an algorithm to do preliminary integration operations automatically by combining propositional representations in a certain way.
However, the initial integration approach is not always satisfactory, addressing temporal relationships is crucial in externalizing narrative texts, as the temporal sequence of fragments doesn't always correspond with the narrative order of the text.
Although \toolName{} enables the creation of multiple text fragments to depict temporal juxtapositions, the influence of sequential relationships in the textual narrative on temporal order is significant, so additional ways of integration will be included in the system.

Clover looks at the storyline view and finds that fragment II, \textcolor{black}{``Dumbledore and McGonagall's discussion of the recent events,"} is placed after fragment I, \textcolor{black}{``The communication between Dursley and the weatherman,"} when in fact the two fragments occur at the same time. So Clover drags fragment II to the same level as Fragment I. \toolName{} automatically reintegrates the fragments and generates a new storyline layout to form the correct mental model.
% Throughout the integration process, Clover can adjust the duration of fragments by adjusting the left and right boundaries of fragments in the storyline view, or freely merge or delete fragments to complete further abstraction of representation.
During this integration stage, Clover has the flexibility to modify fragment durations by adjusting their left and right boundaries, and he can also merge or delete fragments.
% It is worth noting that all operations in the integration stage are presented to the user in a "preview" manner, courtesy of the pre-defined propositional representation integration method, and will be applied only after the user confirms them, in order to ensure the stability of the integration stage (D3). In this way, \toolName{} supports the last step of the selection-organization-storytelling workflow. The final storyline of Clover is shown in Figure~\ref{Fig1}.
Notably, all operations at this stage are presented in a \textcolor{black}{``preview''} mode, based on the pre-defined method. These changes are only finalized upon user confirmation(DC3). In this manner, \toolName{} facilitates the final step in the selection-organization-storytelling workflow. Clover's completed storyline is illustrated in Figure~\ref{Fig1}.

\subsubsection{Re-reading and Adjustment}
% In addition to integrating working memory into long-term memory through our workflow, the rich visual elements of the storyline can be used for rereading and additional revisions.
Beyond embedding working memory into long-term memory through our workflow, the storyline's rich visual elements also support re-reading and further revisions. 
By tightly integrating text and storyline, \toolName{} provides interactive assistance for observation and revision (DC3). When Clover forgot something about a storyline and referred back to it, the storyline served as a visual diagram. He can zoom out for an overview of the entire storyline or zoom in on specific time points.
Below the storyline view is a mini-map for navigating the entire storyline. The pane above the mini-map displays information about the place of the currently displayed storyline.
This bidirectional binding to the storyline enhances Clover's detailed and holistic understanding of the story. 
% At the same time, an axis divided by chapters is placed on top of the mini-map to help Clover have a sense of time. Or, if Clover wishes to view only the plots of particular persons, he can filter the persons by tapping the legend above the storyline view.
Additionally, to focus on a specific person, Clover can filter by the person using the legend above the storyline view.

% Clover revisits the storyline to refresh his memory, when he clicks on a fragment to review the corresponding content in the text, the text view automatically navigates to the original text. 
% During the entire rereading stage, Clover can utilize his newly generated understanding of the text to dynamically adjust the storyline layout.
% Clover is also allowed to add or remove persons through the person entity alternatives in fragment view or to change the content of other fragment elements through direct text editing. 
% These interactions support rereading by allowing Clover to switch between text view (the text), fragment view (the fragment) and storyline (the diagram). The views are interdependent, and changes to one view will result in synchronized changes to the others.

% If Clover wants to share his favorite stories with others, he can use our generated storyline directly when he shares various narrative texts through this series of steps. To make this even better, Clover can activate the review mode in the toolbar above the text view, which hides the entire storyline, and when he clicks on the storyline view step by step, the storyline will be displayed in chapter, page, or custom format (similar to the way a PowerPoint presentation is displayed), thus focusing the observer's attention more. \toolName{} is also working on additional extensions of the review method (e.g., redrawing the storyline, implementing spaced learning, etc.). We believe this will be very beneficial for teaching and learning.

\begin{figure*}[t]
    \centering
    \includegraphics[width=\linewidth]{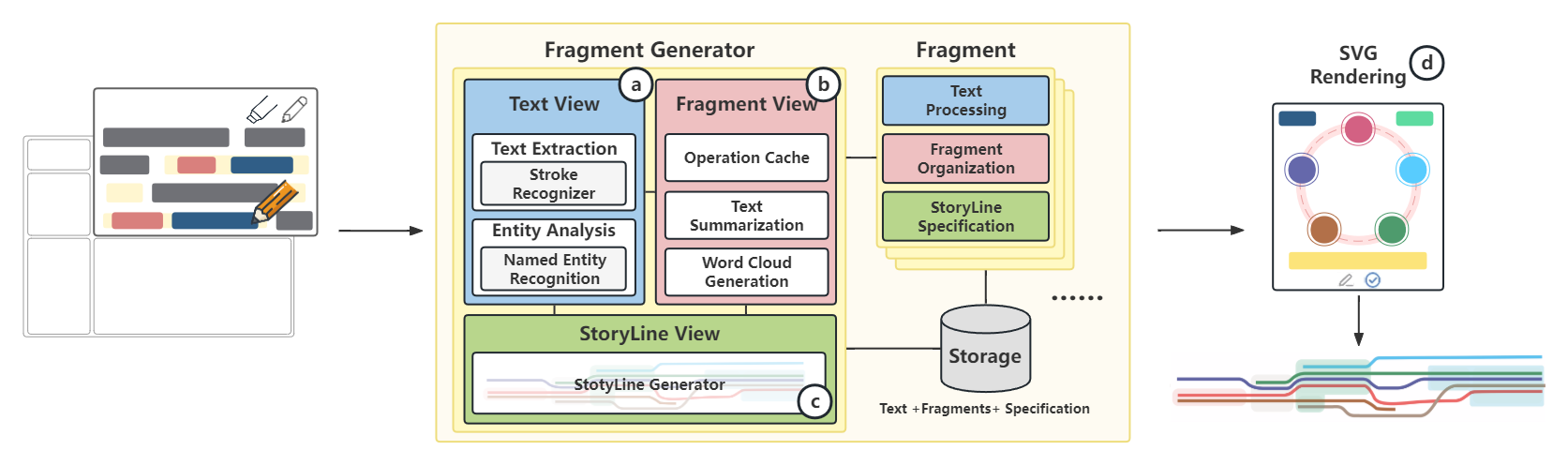}
    \caption{\toolName{} system architecture: (a) text view integrates stroke annotation and GPT-based visual hints to extract entities, (b) fragment view records user operations and generates text summarization and word cloud. (c) storyline view creates storytelling specifications, and (d) SVG renderer updates the diagram canvas.}
    \label{Fig5}
    \vspace{-6mm}
\end{figure*}

\vspace{-4mm}
\subsection{User Interface and System Implementation}

The system's interface comprises four main views. 
In the configuration view (Figure~\ref{Fig1}a), users can import text and view statistics, and adjust parameters for entity recognition and text summarization.
The fragment view (Figure~\ref{Fig1}c) facilitates operations like viewing, modifying, and deleting fragment elements.
The text view (Figure~\ref{Fig1}b) allows for highlight and stroke annotations on texts via PC or tablet, accessible through the top toolbar,
and the storyline view (Figure~\ref{Fig1}d) presents a zoomable, scrollable canvas for storyline interaction, with a miniature map for focused chart viewing. \toolName{} is a web-based client-server application that utilizes Flask to analyze and store text, fragments, and SVG specifications. The client side developed in HTML and JavaScript, manages annotations, entity imports, and renders SVG elements.

\subsubsection{Text View}
The module's main function is to extract entities selected or annotated by end users, consisting of a text extraction and an entity analysis module. 
The text extractor analyzes user stroke annotations to extract corresponding entity text content. The stroke recognizer in the text extraction is implemented using the NDollar recognizer \cite{anthony2010lightweight}. By using pre-set stroke annotations as a dataset, each data contains an array of points with x and y coordinates, as well as specific attributes used for recognition. \textcolor{black}{The training data is continuously updated to improve accuracy based on user marking patterns.}

Recognition hinges on calculating distances and positional relationships (top, middle, bottom) between text and annotations. Additionally, the module assesses historical annotation positions to ascertain if the marked content constitutes multi-line text linked to the same entity.

% TODO：lei Wang
Following the extraction of entity text, it becomes imperative to discern implicit text entities within sentences. \textcolor{black}{We employ GPT-3.5 in combination with the Chain-of-Thought model proposed by Wei et al.\cite{wei2022chain}, using a series of prompts to guide GPT in producing inference-driven outcomes.} The prompts exhibit three key features: a chained structure, knowledge load balancing, and iterative generation.

\noindent
\textbf{Chain-of-Thought model.}
We defined the model's function using natural language, as illustrated by, ``I anticipate the model to assume the role of a proficient data analyst adept at extracting information from textual data.'' \textcolor{black}{Subsequently, we devised a set of chain rules to guide the model in extracting relevant entities from the text. The chain rules are as follows:}
% \begin{itemize}
%     \setlength{\itemsep}{0pt}
% 	\setlength{\parsep}{0pt}
% 	\setlength{\parskip}{0pt}
%     \item
%    \textcolor{black}{ \textbf{Read the Text}: Conduct a thorough reading of the given text.}
%     \item
%   \textcolor{black}{\textbf{Identify Entities}: Based on the text’s theme and context, mark key entities within the text.}
%     \item 
%     \textcolor{black}{\textbf{Format Output}: Convert the identified entities into JSON format to save the extracted information.}
% \end{itemize}

\noindent
\textbf{Multi-knowledge sources.}
\textcolor{black}{In our prompt design, we integrate external knowledge to enhance the model's comprehension. When faced with limited user inputs, we guide the model to rely more on its training data for accurate entity extraction. We've implemented a customizable trust threshold, allowing users to set the confidence level for balancing external knowledge and user-provided information.}

\noindent
\textbf{Iterative generation rule. }
Jha et al.\cite{jha2023dehallucinating} demonstrated the viability of the iterative generation mechanism in guiding large language models (LLMs) to converge toward the correct final output through multiple rounds of refinement. \textcolor{black}{We've incorporated this iterative approach into our entity extraction process. In each iteration, users provide a marked list of entities, which the model uses to refine previously ambiguous entity names.}

\subsubsection{Fragment View}
Once the final entity is acquired, the fragment organization module arranges it further into a fragment. This module includes an operation cache, a text summarization model, and a word cloud generator for extracting fragment keywords. \textcolor{black}{The operation cache maintains fragment-level caches that log entity modifications and operation histories, prioritizing modified entities in future fragment creation.}
% TODO：lei Wang
If the user does not select the corresponding event element, the module will initiate a concise analysis of the content using GPT-3.5. This analysis follows the methodology outlined in Section 5.1, specifically employing the Chain-of-Thought model for the generation of text summarization.

Enriching the visual expression of the storyline by selecting keywords in the word cloud is our important design principle.
\textcolor{black}{Our approach to word cloud generation builds upon the foundational work of ReCloud \cite{wang2020recloud}. We utilize advanced natural language processing techniques, including deep learning-based syntax dependency analysis, to construct a comprehensive semantic graph of the fragment's content.} User annotations significantly contribute to the word cloud's content. We assign varying weights to different levels of annotated text, including all text, highlighted portions, and annotated or recognized entities within the fragment. By incrementally adjusting these weights in our system, we can create a word cloud that more effectively encapsulates the essence of the fragment.

\subsubsection{StoryLine View}
Once fragment content is organized, the module integrates multiple fragments through user actions such as creation, modification, and deletion to form our pre-set proposition representation. \textcolor{black}{We extend Tanahashi's\cite {tanahashi2012design} storyline data specification with additional information for visual design and generalize fragments into a consistent, persistent data structure.} The StoryFlow\cite{liu2013storyflow} algorithm is employed to visualize different fragments into a storyline layout. \textcolor{black}{We have enhanced the interactive capabilities for viewing and editing the storyline using D3.js, incorporating features such as zooming, panning, and direct manipulation of storyline elements.}

\section{EVALUATION}
To evaluate \toolName{}, our aim was to assess whether it can create a satisfactory storyline from narrative texts. To achieve this, we conducted a case study and user interviews to demonstrate the effectiveness of our approach.

\subsection{Case Study}
\textcolor{black}{In this section, we describe how \toolName{} can be used in practice to construct a storyline and externalize knowledge from raw text through two case studies with two target users (U4, \textcolor{black}{U8}) from our formative study. Each participant received a compensation of \$10 for their participation.}

\begin{figure*}[t]
    \centering
    \includegraphics[width=0.8\linewidth]{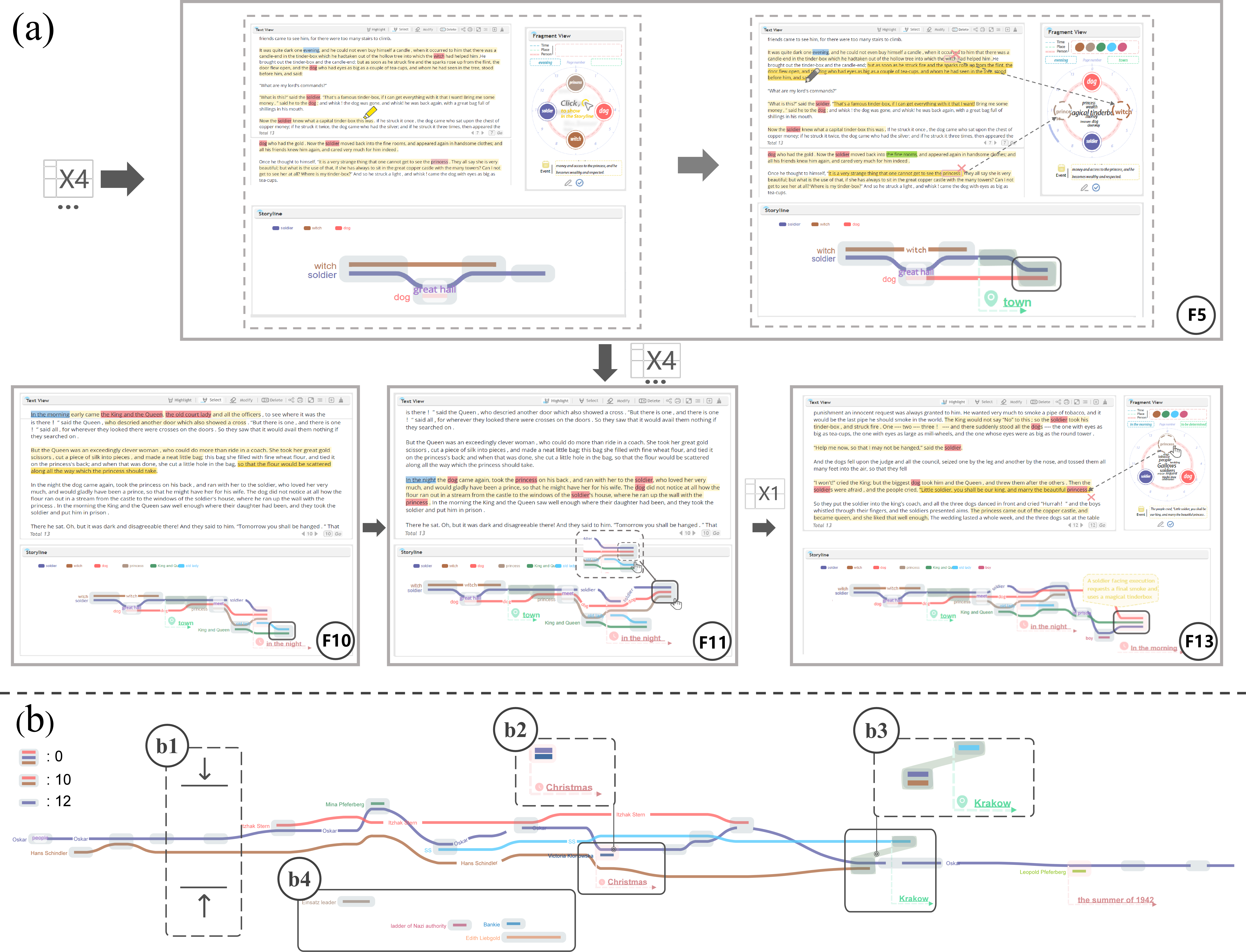}
    \caption{Two cases of using \toolName{} to construct storylines for the novel \textcolor{black}{``The Little Match Girl''} and the script of \textcolor{black}{``Schindler's List''} are presented. The entire workflow follows the selection-organization-storytelling method. In Figure F5, U4 employed the highlight-select method, initially constructing a fragment and filtering characters by deleting them from the text. Figure F11 demonstrates how U4 adjusted a fragment that was misclassified as occurring at the same time. Figure F13 illustrates the interconnectivity of all views, where actions in the fragment view are reflected in the text view. Figure b1 displays the relatively flat storyline constructed from the script. In Figure b2, crucial time points are highlighted in the storyline. In Figure b3, two events at the same location are emphasized, indicating their corresponding places. Figure b4 showcases the script's construction of storylines with multiple single-character fragments.}
    \label{FigCase}
    \vspace{-6mm}
\end{figure*}

\noindent
\textbf{Matchstick Girl: Constructing a Storyline from Fiction.}
U4 is curious about our system and eager to experience \toolName{} to discern the differences between our tool and traditional pen-and-paper storyline construction.

U4 was curious about abstracting the knowledge externalization process from narrative texts into a cohesive workflow. \textcolor{black}{``Generally, users navigate through fragmented construction processes, which is inefficient. A unified workflow would be intriguing"}. Considering this, U4 began by examining how the system captures various entities in the text. He used the \textit{Highlight} tool to mark the part he was interested in and, upon selection, he noticed that entities automatically appear in the highlighted text (Figure~\ref{FigCase}-F5). He found this interaction approach novel and convenient, eliminating the need to identify and select entities repeatedly.
U4 then discovered that a \textcolor{black}{``location"} entity is not recognized. He used the \textit{Selece} tool to annotate the entity \textcolor{black}{``the fine room"} and confirmed its type through the entity type selector. ``This is obviously more convenient than pen-and-paper. Combining automatic entity recognition and stroke annotation allows me to build fragments quickly.'' He also noticed that the person entities \textcolor{black}{``prince"} and \textcolor{black}{``witch"} are misidentified. The \textcolor{black}{``Delete"} function enables him to quickly remove these incorrect entities from the text.

As U4 naturally shifted his focus to organizing fragments, he refined them and selected keywords from a word cloud. He noticed the storyline being iteratively built after applying these fragments. \textcolor{black}{``This interaction speeds up the overall externalization process and feels natural."}
After successfully iterating a few fragments, U4 observed a temporal discrepancy in the storyline: different time-segmented fragments appear as the same time segment (Figure~\ref{FigCase}-F11). He found that all fragments and visual hints in the storyline view are draggable and adjustable. He can easily move and resize fragments. Moreover, he also discovered a key feature: changes in one view of the system automatically update all other views (Figure~\ref{FigCase}-F12). This inter-view binding further simplifies the operation.

\begin{figure*}[t]
    \centering
    \includegraphics[width=\linewidth]{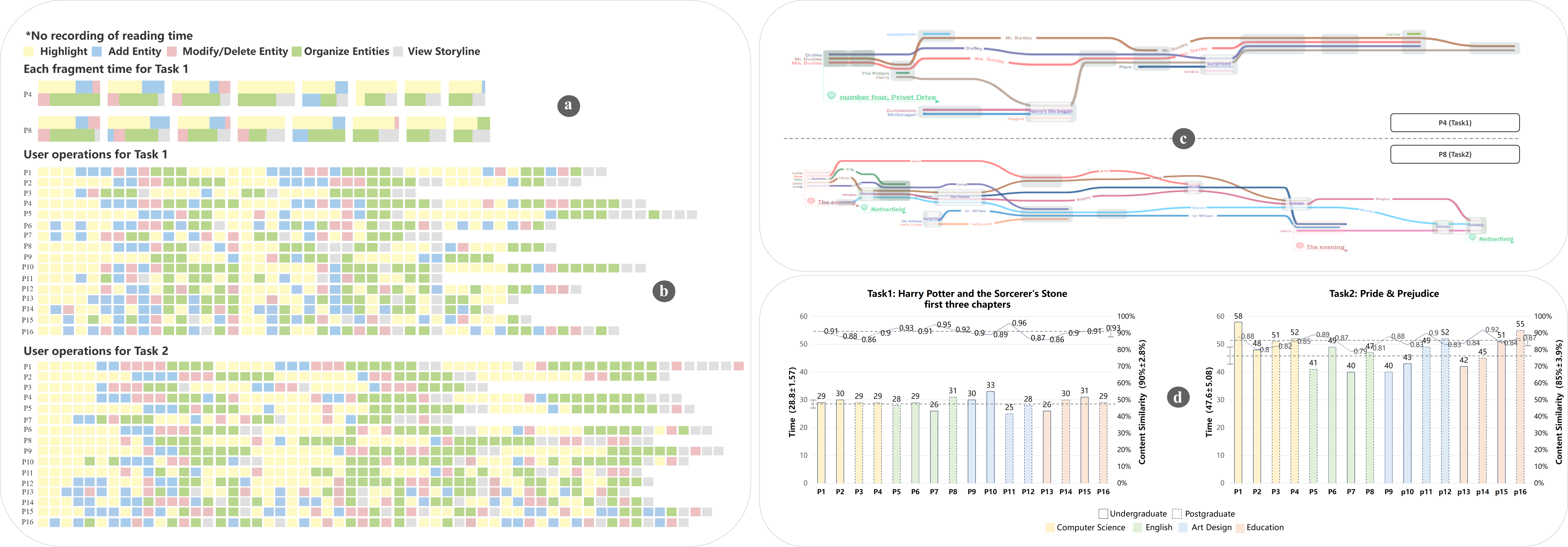}
    \caption{\textcolor{black}{User interview Results: (a) the first seven fragments of each operation time statistics for Task 1 “Harry Potter and the Sorcerer's Stone first three chapters”, where the block size indicates the length of time spent. (b) StoryExplorer log-data visualization showing different actions. (c) Storyline created by U4 for task 1 \textcolor{black}{``Harry Potter and the Sorcerer's Stone first three chapters"} and \textcolor{black}{U8} for task 2 \textcolor{black}{``Pride \& Prejudice"}. (d) Similarity of final storyline and completion time statistics for both tasks. As the complexity of the text increases, both the time costs and content similarity are significantly affected.}}
    \label{FigCombine}
    \vspace{-6mm}
\end{figure*}

\noindent
\textbf{Schindler's List: Crafting a Narrative from Script.}
The structural variances between script text and novel text are substantial. Scripts, prioritizing dialogues, may pose challenges for non-professionals in the film and television industry. To address this, we engaged a targeted user (\textcolor{black}{U8}) with a background in film and television to evaluate the effectiveness of our tool in externalizing knowledge during script text reading.

Throughout the annotation process, \textcolor{black}{U8} observed the automatic entity recognition in highlighted text, expressing appreciation for its assistance in deciphering complex dialogue scenarios. Additionally, \textcolor{black}{U8} acknowledged the system's proficiency in automatically identifying temporal and spatial elements in highlighted text, particularly beneficial in dialogue-intensive scripts where readers often struggle to discern the time and location of events within dense dialogue fragments. Transitioning from text annotation to storyline construction, \textcolor{black}{U8} found that \toolName{} intelligently organizes multiple annotated fragments into a cohesive storyline. Notably, the tool highlights story fragments occurring in the same location, a functionality highly praised by \textcolor{black}{U8} for its assistance in the comprehensive observation of interrelated plots(Figure~\ref{FigCase}-b3).

\textcolor{black}{We observed notable differences in storylines created from scripts versus novels. The scene-based narrative structure of scripts, featuring relatively fewer characters in each scene, tends to produce a flatter storyline(Figure~\ref{FigCase}-b1). In contrast, novels often employ multiple scene transitions on the same timeline, resulting in more diverse and dynamic narratives. Furthermore, in the storyline created by \textcolor{black}{U8}, we noted multiple fragments featuring a single character(Figure~\ref{FigCase}-b4). \textcolor{black}{U8} mentioned this reflects the monologue in the script text. Regarding the narrative intersections and twists, \textcolor{black}{U8} recognized its exceptional narrative expressive capabilities. To enhance the clarity of narrative understanding, we introduced time points in the storyline(Figure~\ref{FigCase}-b2). This improvement emphasizes critical fragments and provides a clear timeline for intersecting plot elements.}

\subsection{User Interviews}
We conducted in-depth user interviews with 16 target users to evaluate the performance of \toolName{} for given tasks.

\subsubsection{Study Design}
Our user study is divided into three parts. 
Part 1 (Understanding fragment organization approach) evaluates whether people can quickly understand the interaction mode of \toolName{}. 
Part 2 (Enhancing efficiency in generating storyline) evaluates whether \toolName{} can efficiently and effectively generate a complete storyline based on time and content similarity. 
Part 3 (Reducing cognitive load through workflow) further validates that \toolName{} can help users reduce cognitive load and form more coherent mental models while generating good narrative structures, based on cognitive load theory.

\textbf{Participants.} 
\textcolor{black}{The study included 16 participants (10 female, 6 male; age M=22.0 years, range 20-28; P1-P16) from a research university. All participants had normal or corrected-to-normal vision and were majoring in Computer Science, English, Art Design, or Education. Most participants reported prior experience with document annotation tools (e.g., Adobe Acrobat \cite{AdobeAcrobatPro}, TexSketch \cite{subramonyam2020texsketch}).}

\textbf{Apparatus.}
Our work was conducted on a PC platform. We utilized a 24-inch LCD monitor with a resolution of 1920 × 1080. 
Throughout the user interview, all participants were physically present offline.

\textbf{Tasks.}
We prepared two tasks for each of the three evaluation parts, one for a text with relatively clear entity relationships (Task 1, Harry Potter and the Sorcerer's Stone first three chapters) and one for a text with relatively complex entity relationships (Task 2, Pride \& Prejudice). The experiments included learning and using the basic interaction of \toolName{}, completing the required storyline for the task, and completing a cognitive load survey after the experiment. For each of these tasks, the expression and complexity of subsequent tasks gradually increased.

\textbf{Procedure.}
\textcolor{black}{The study began with research objectives, procedures, and a brief overview of \toolName{}, including storyline creation instructions.} Research assistants introduced the workflow by implementing our formative study case, ``The Tinder-Box." Once the participants indicated familiarity, they started with our customized first task. We recorded the participants' interaction patterns, including partial (time for each operation) and global (time for building the storyline). We also rated the content of the produced storyline to evaluate Parts 1 and 2. \textcolor{black}{Subsequently, we asked the participants to complete the Cognitive Load Survey (CLS)\cite{leppink2013development} to measure their subjective cognitive load during the task, and took a ten-minute break before starting the next task.}

After completing all the tasks, the participants filled out a usability questionnaire \cite{lund2001measuring}, where they were asked to rate \toolName{} on a 7-point scale, with 1 being \textcolor{black}{``strongly disagree"} and 7 being \textcolor{black}{``strongly agree"}. The study concluded with semi-structured interviews to collect feedback on the \toolName{} experience. In addition, we captured all the iterations of the entire process along with \toolName{} log data for further analysis to evaluate Part 3. On average, the entire session lasted approximately 90 minutes (30 minutes for the first task, 50 minutes for the second task, and a 10-minute break).

\subsubsection{Interview Results}

% % \begin{figure}[b]
% %     \centering
% %     \includegraphics[width=\linewidth]{figs/FigCaseRes.png}
% %     \caption{Evaluation Results: Storyline created by P4 for task 1 "Harry Potter and the Sorcerer's Stone first three chapters" and \textcolor{black}{P8} for task 2 "Pride \& Prejudice".}
% %     \label{FigCaseRes}
% % \end{figure}

% % \begin{figure}[b]
% %     \centering
% %     \includegraphics[width=\linewidth]{figs/Fig6.png}
% %     \caption{Evaluation Results: (a) the first seven fragments of each operation time statistics for Task 1 “Harry Potter and the Sorcerer's Stone first three chapters”, where the block size indicates the length of time spent. (b) StoryExplorer log-data visualization showing different actions.}
% %     \label{Fig7}
% % \end{figure}

\textbf{Understanding narrative text AD workflow.}
\textcolor{black}{Analysis of the results (Figure~\ref{FigCombine}c) and participant interviews revealed the effectiveness of the two-pass reading workflow for narrative text organization. The majority of participants utilized the highlighting feature to construct fragments prior to organization, indicating that this approach facilitated their preliminary mental model construction. As \textcolor{black}{P8} noted, this two-pass approach aligned well with natural reading patterns while effectively conveying the narrative structure. However, one notable exception was P10, who chose to manipulate text entities during the highlighting phase, resulting in expedited construction but lower content quality scores.}

\textcolor{black}{The organization phase revealed particularly positive responses to the GPT-based entity identification system.} P1 reported, \textcolor{black}{``}When I finished highlighting the text and started the second pass, I found that the required character and location information had already been annotated." Participants were generally satisfied with the GPT-based annotation method, although P2 felt that \textcolor{black}{``}the automatically identified entities are not always satisfactory, and I need to make additional adjustments to the characters or places. But once modified, the same content will change together, which is very convenient."

\begin{figure}[t]
    \centering
    \includegraphics[width=\linewidth]{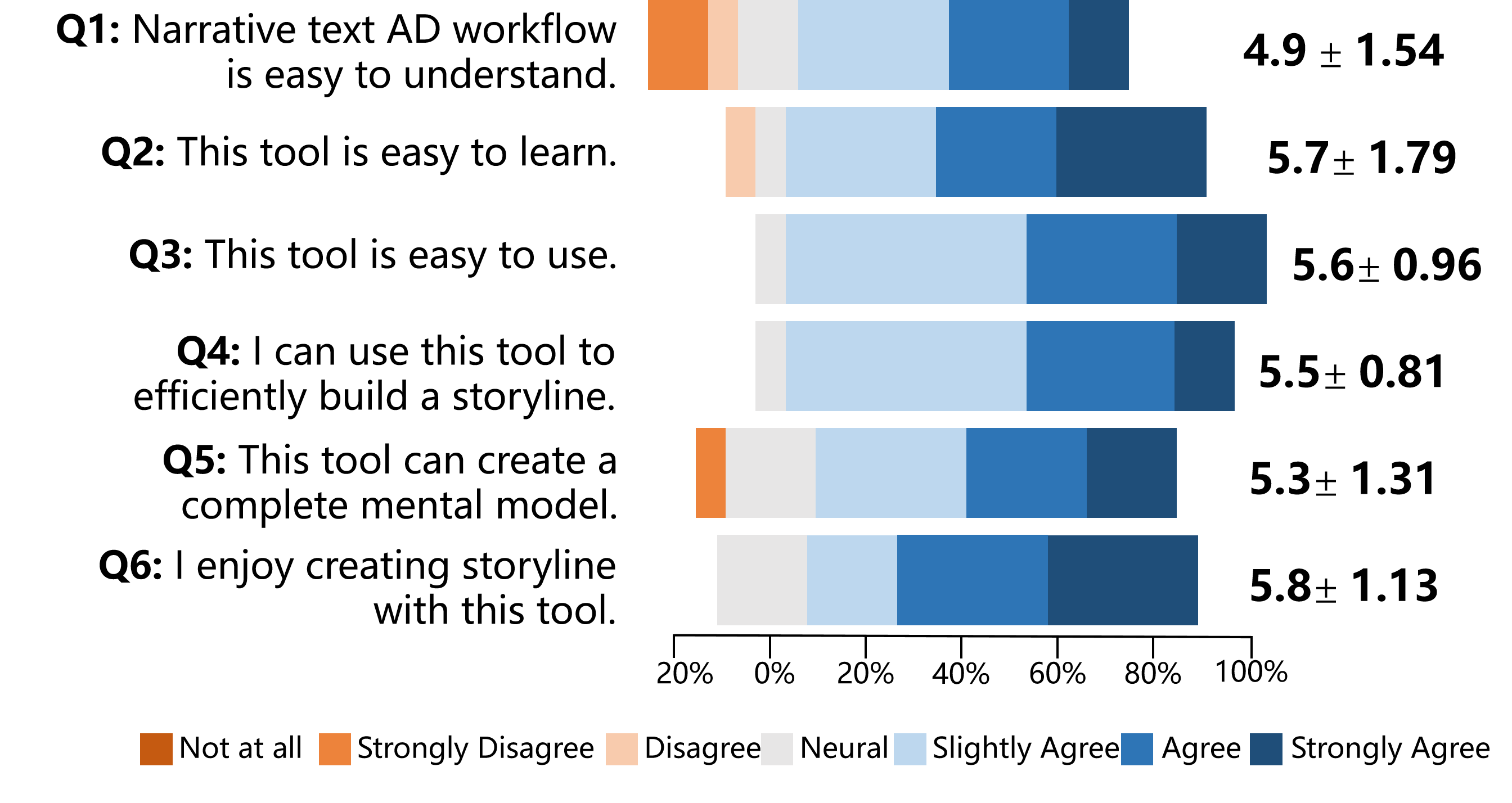}
    \caption{\textcolor{black}{Evaluation Results: Results of the six questions in the usability questionnaire: The rightmost column denotes Mean ± SD, and the score is based on a seven-point scale.}}
    \label{FigQuestionaire}
    \vspace{-3mm}
\end{figure}

\textcolor{black}{Surprisingly, the storytelling phase demonstrated high performance across all participants, with users quickly adapting to the storyline view's interface design. As one participant noted ``The interface's similarity to existing software enabled immediate productivity." This qualitative feedback was supported by quantitative data showing a significant decrease in per-fragment operation time during reading progression (Figure~\ref{FigCombine}a).} \textcolor{black}{Post-study questionnaire results (Figure~\ref{FigQuestionaire}) further validated the system's effectiveness, with participants rating the narrative text AD workflow as comprehensible (M=4.9), learnable (M=5.7), and usable (M=5.6).}

\textbf{Enhancing efficiency in generating storyline.}
\textcolor{black}{System log analysis demonstrated the efficiency of our GPT-based entity annotation approach (Figure~\ref{FigCombine}b). User interaction patterns showed participants focused primarily on core tasks - text highlighting and fragment organization, with minimal time spent on entity operations. The cognitive load of entity manipulation decreased as users became familiar with the system.}

\textcolor{black}{Task completion analysis (Figure~\ref{FigCombine}d) showed mean completion times of 28.8 minutes (SD=1.57) for Task1 and 47.6 minutes (SD=5.08) for Task2, excluding one outlier (P1). The relatively small standard deviation suggests consistent performance across users.} \textcolor{black}{We drew the correct storyline in advance as the ground truth. The ground truth storyline was collaboratively created by recruited graduate students in literature and co-authors familiar with the system operation, and then all authors checked and finalized it. }\textcolor{black}{We compared participants' final storylines against a ground truth storyline created by literature graduate students and validated by all authors. The average content similarity was 90\% (Task1) and 85\% (Task2). This high accuracy, combined with post-task confidence ratings (M=5.5), indicates effective user performance with the system.}
% \begin{figure*}[t]
%     \centering
%     \includegraphics[width=0.7\linewidth]{figs/FigTime&Cs.png}
%     \caption{Evaluation Results: Similarity of final storyline and completion time statistics for both tasks. As the complexity of the text increases, both the time costs and content similarity are significantly affected.}
%     \label{FigTime&Cs}
% \end{figure*}

\textbf{Reducing cognitive load through workflow.}
\textcolor{black}{Participants reported high engagement with \toolName{}, emphasizing its effectiveness in text organization and comprehension enhancement. One participant (P2) notably remarked: \textcolor{black}{``}If I had used this tool before, I could have done better on my exam." This positive user experience was reflected in the 7-point usability questionnaire (Figure~\ref{FigQuestionaire}), with participants rating the tool's ability to generate complete mental models (M=5.3) and storyline creation enjoyment (M=5.8) favorably.}

\begin{figure}[h]
    \centering
    \includegraphics[width=\linewidth]{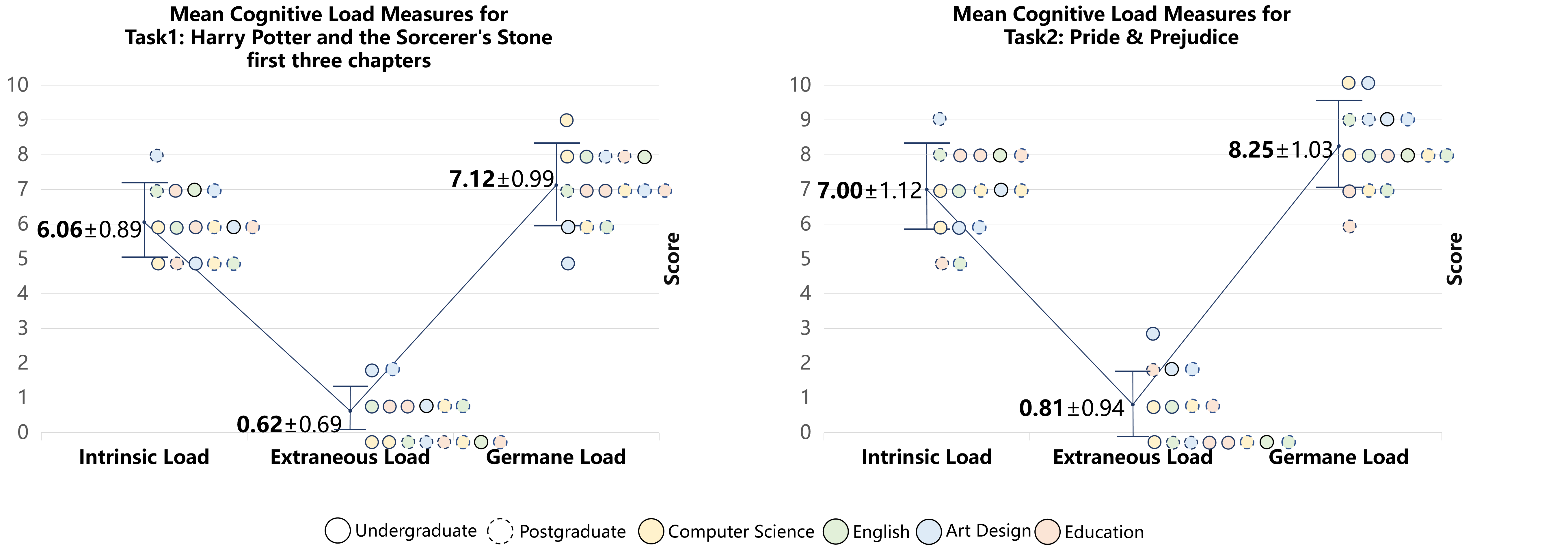}
    \caption{\textcolor{black}{Results from cognitive load survey (IL=intrinsic load, EL=extraneous load, GL=germane load)}.}
    \label{FigCLS}
    \vspace{-6mm}
\end{figure}

\textcolor{black}{Cognitive Load Scale (CLS) assessment revealed favorable learning conditions across both tasks. The CLS questionnaire uses a ten-point rating scale to measure (1)the intrinsic load (IL) corresponding to the difficulty of the reading material, (2) the extraneous load (EL) of the tool and reading intervention, and (3) the germane load (GL) of participants’ self-perceived learning. Participants reported high intrinsic load (Task1: M=6.06, SD=0.89; Task2: M=7.00, SD=1.12) and notably low extraneous load (Task1: M=0.62, SD=0.69; Task2: M=0.81, SD=0.94), indicating that while the reading material was appropriately challenging, the tool itself posed minimal cognitive burden. Self-perceived learning scores were particularly high (Task1: M=7.12, SD=0.99; Task2: M=8.25, SD=1.03), suggesting effective knowledge acquisition (Figure~\ref{FigCLS}).}

\section{DISCUSSION}
\textcolor{black}{We first summarize lessons learned during the development of \toolName{}. Then, we discuss the limitations of \toolName{}. Finally, we outline future work to address these limitations and expand the capabilities of \toolName{}.}

% learn and lessons  align with narrative orders
\vspace{-2mm}
\subsection{Lessons}
% \textbf{Supporting Better Entity Recognition.}
% After using \toolName{}, some participants (P2, \textcolor{black}{P8}, P10) were not completely satisfied with entity recognition using NLP, especially in Task2 where many entities needed to be manually added or modified. Additionally, disregarding name prefixes resulted in multiple persons being identified as one, which is concerning. However, they also acknowledged that the current recognition results were helpful for quick entity selection and fragment organization. In the future, more advanced techniques in named entity recognition should be employed to improve accuracy.

\noindent
\textbf{Enhancing Temporal Juxtaposition.}
\textcolor{black}{In the storyline visualization field, having correct temporal juxtaposition is crucial for visual specification. However, in the current design, the temporal position of fragments depends on their appearance in the text or manual modification in the storyline, which is not in line with our design considerations of rapid iteration for the storyline.
To address the current limitation of fragment positioning, we could better utilize text corpus information and complete fragment alignment through entity information such as time or place.}

\noindent
\textbf{Intelligent Recommendation Fragment.}
\textcolor{black}{Several participants (P2, P4, P15) mentioned during interviews that they hoped for automatic extraction of fragments from the text, with manual intervention only during modification for convenience. Our computer science background students (P2, P3) also suggested existing event extraction algorithms may help achieve this. While existing event extraction algorithms have limitations, recent research suggests rule-based approaches or more universal event extraction algorithms could help solve this problem.}
\vspace{-8mm}
\subsection{Limitations}

\noindent
\textbf{Performance Issues.}
\textcolor{black}{The system design has performance bottlenecks, particularly in GPT model interface calls. Future improvements include preprocessing imported text, reducing calls, and preloading entities to enhance user experience.}

\noindent
\textbf{Evaluation Limitations.}
\textcolor{black}{Our study participants were limited to students with similar age ranges and academic backgrounds. Future evaluations should include a more diverse participant pool to account for different active reading strategies based on age, teaching methods, and research environments.}

\vspace{-4mm}
\textcolor{black}{\subsection{Future Work}}

\noindent
\textcolor{black}{\textbf{Expanding Text Type Applicability.}
While \toolName{} currently focuses on novels and scripts, future research will investigate its potential for a broader range of text types, such as news articles and conference abstracts. This expansion will help determine the tool's scalability across various narrative structures and information densities.}

\noindent
\textcolor{black}{\textbf{Long-term Memory Assessment.}
Although \toolName{} aims to facilitate the transition from working memory to long-term memory through storyline extraction, the current study does not evaluate its effectiveness in long-term memory retention. Future research will include methods to assess \toolName{}'s impact on long-term memory through longitudinal studies or comparative analyses (e.g., delayed recall tests at various intervals and evaluation of story element retention patterns) versus traditional reading methods.}

\noindent
\textcolor{black}{\textbf{Collaborative Reading Platform.}
We plan to explore how \toolName{} can enable collaborative active reading through social platforms (e.g., nb \cite{zyto2012successful}), integrating features for shared annotations and collective storyline construction.}

 \noindent
\textcolor{black}{\textbf{Neutral Evaluation Metrics.}
The positively framed statements in the questionnaire, like ``the workflow is easy to learn," may bias participants toward higher ratings\cite{yang2018does}. To avoid framing effects on participants' subjective ratings, we plan to use more neutral statements in future studies.}

\section{CONCLUSION}
Based on models for externalizing textual narratives, we proposed a workflow for storyline generation of narrative text. Through formative study, we identified key design considerations in extracting narrative structures. The result is \toolName{}, a system in which readers can select potential entities and construct story fragments using stroke annotation and GPT-based visual hints. The selection-organization-storytelling workflow helps readers convert entity content at the text level into fragment level and integrate them through the storyline. GPT-based visual hints help focus readers' attention, while the visually rich storyline improves the formation of lasting, coherent mental models. Through case studies and user interviews, we evaluated the comprehensibility of the workflow, as well as the learnability and usability of \toolName{}.
% The results show our approach provides readers with the opportunity to learn narrative text through a carefully designed workflow, while achieving good cognitive load performance.
The results indicate that we offer an effective workflow for users to externalize knowledge from narrative texts, while also achieving good cognitive load performance.

\vspace{-2mm}
% \begin{acks}
% We would like to thank the reviewers for their thoughtful comments. This work was supported in part by the National Natural Science Foundation of China (No.62277013, No.62177040, No.62132017), Zhejiang Lab Open Research Project (No.K2022KG0AB01), National Statistical Science Research Project (No.2022LY099), Zhejiang Provincial Science and Technology Program in China (No.2021C03137), Public Welfare Plan Research Project of Zhejiang Provincial Science and Technology Department (No.LTGG23H260003), and Zhejiang Statistical Science Research Project.

% \end{acks}

% %-------------------------------------------------------------------------
% bibtex
% \bibliographystyle{eg-alpha-doi}  
\bibliography{egbibsample}        
\bibliographystyle{IEEEtran}

% % biblatex with biber
% % \printbibliography                

% %-------------------------------------------------------------------------
% %Color tables are no longer required for purely electronic publications.
% % \newpage
% % 
% % 
% % \begin{figure*}[tbp]
% %   \centering
% %   \mbox{} \hfill
% %   % the following command controls the width of the embedded PS file
% %   % (relative to the width of the current column)
% %   \includegraphics[width=.3\linewidth]{sampleFig}
% %   % replacing the above command with the one below will explicitly set
% %   % the bounding box of the PS figure to the rectangle (xl,yl),(xh,yh).
% %   % It will also prevent LaTeX from reading the PS file to determine
% %   % the bounding box (i.e., it will speed up the compilation process)
% %   % \includegraphics[width=.3\linewidth, bb=39 696 126 756]{sampleFig}
% %   \hfill
% %   \includegraphics[width=.3\linewidth]{sampleFig}
% %   \hfill \mbox{}
% %   \caption{\label{fig:ex3}%
% %            For publications with color tables (i.e., publications not offering
% %            color throughout the paper) please \textbf{observe}: 
% %            for the printed version -- and ONLY for the printed
% %            version -- color figures have to be placed in the last page.
% %            \newline
% %            For the electronic version, which will be converted to PDF before
% %            making it available electronically, the color images should be
% %            embedded within the document. Optionally, other multimedia
% %            material may be attached to the electronic version. }
% % \end{figure*}

\end{document}